# Atomic-Scale Heterogeneity of Hydrogen in Metal Hydrides Revealed by Electron Ptychography

Pengcheng Li[1,2]†, Chenglin Pua[2]†, Zehao Dong[3], Zhengxiong Su[4], Tao Liu[2], Chao Cai[5], Huahai Shen[6], Lin Gu[2]*, Zhen Chen[1,7]*

**Affiliations:**

[1]Beijing National Laboratory for Condensed Matter Physics, Institute of Physics, Chinese Academy of Sciences, Beijing, 100190, China.
[2]Beijing National Center for Electron Microscopy and Laboratory of Advanced Materials, School of Materials Science and Engineering, Tsinghua University, Beijing, 100084, China.
[3]State Key Laboratory of Low Dimensional Quantum Physics, Department of Physics, Tsinghua University, Beijing, 100084, China.
[4]School of Nuclear Science and Technology, Xi'an Jiaotong University, Xi'an, 710049, China.
[5]Department of Materials Science and Engineering, Southern University of Science and Technology, Shenzhen, 518055, China.
[6]Institute of Nuclear Physics and Chemistry, China Academy of Engineering Physics, Mianyang, 621900, China
[7]School of Physical Sciences, University of Chinese Academy of Sciences, Beijing 100049, China.

*Corresponding author. Email: lingu@mail.tsinghua.edu.cn; zhen.chen@iphy.ac.cn.

†These authors contributed equally to this work.

**Abstract:** Hydrogen plays critical roles in materials science, particularly for advancing technologies in hydrogen storage and phase manipulation, while also posing challenges like hydrogen embrittlement. Understanding its behavior, vital for improving material properties, requires precise determination of atomic-scale distribution—a persistent challenge due to hydrogen's weak electron scattering and high mobility, as well as the limitations of conventional transmission electron microscopy. We demonstrate that multislice electron ptychography (MEP) overcomes these constraints through three key advances: exceptional sensitivity for hydrogen occupancy, three-dimensional quantification, and picometer-level precision in atomic positioning. Experimentally, MEP resolves heterogeneous hydrogen distributions and quantifies hydrogen-induced lattice displacements with picometer precision in multi-principal-element alloy hydrides. This work demonstrates MEP as a transformative method for directly probing hydrogen atoms in solids, unlocking fundamental understanding of hydrogen's impact on material properties.

**Main Text:**

Hydrogen atoms, due to their small size, readily incorporate into interstitial sites in alloys and oxides (*1-4*), leading to lattice expansion (*3, 4*), phase transformations (*5-7*), and changes in electronic structure (*2, 3*). These effects play critical roles in phenomena such as hydrogen embrittlement (*8-10*), hydrogen storage (*5-7*), catalysis (*3, 4*), and insulator-metal transitions (*1*). Understanding these behaviors requires precise determination of the occupation sites, atomic positions, and spatial distribution of hydrogen atoms. The structural characterization of hydrogen atoms routinely relies on spatially averaged spectroscopy (*11, 12*) and diffraction methods (*13-15*); however, these approaches inherently fail to capture the inhomogeneous and aperiodic distribution of hydrogen. Direct imaging of hydrogen in real space at the atomic scale remains challenging because of its lowest atomic number and weakest scattering cross section. Recent advances in aberration-corrected optics and phase contrast imaging have significantly improved capabilities of transmission electron microscopy (TEM) (*16-20*), enabling the detection of light elements such as oxygen (*19, 20*), nitrogen (*18-20*), and lithium (*21*). Hydrogen columns have also been resolved using phase contrast techniques such as annular bright field (ABF) (*22*) and integrated differential phase contrast (iDPC) (*23*). However, the image contrast in these methods can be destructively affected by many experimental parameters—including sample thickness, orientation, surface amorphization, and imaging conditions—fundamentally restricting hydrogen observability (*20, 24-26*). Moreover, ABF and iDPC provide only two-dimensional projections and cannot resolve three-dimensional lattice distortions or chemical inhomogeneities, which are common yet critical for understanding hydrogen's impact on material properties. These limitations are particularly detrimental in the complicated multi-principal-element alloys (MPEAs), which have attracted increasing attention for their exceptional mechanical strength and hydrogen storage capabilities (*27-30*).

A promising approach is multislice electron ptychography (MEP), a recently developed phase-retrieval method that can reconstruct a sample's 3D atomic structure from a single dataset (*31, 32*). Unlike conventional scanning transmission electron microscopy (STEM), MEP solves the inverse multiple-scattering problem, reaching sub-angstrom lateral resolution and sub-nanometer depth resolution (*33*), that standard STEM cannot achieve. In addition, the linear phase response of MEP enables quantitative measurement of element concentrations (*34*). Its phase contrast makes it possible to see both heavy and light elements (*31, 34-36*), and great successes have been demonstrated recently in imaging light elements, especially oxygen (*34, 35*). However, direct three-dimensional visualization of hydrogen atoms has not been achieved, mainly due to their extremely low scattering cross-section and sensitivity to electron irradiation. Here, we demonstrate atomic-resolution imaging of hydrogen sublattices in transition metal hydride and MPEA hydride using energy-filtered multislice electron ptychography (EFMEP), an improved version of MEP. Importantly, we show that EFMEP not only reveals atomic positions with high clarity but also uniquely enables quantitative measurement of hydrogen content. By combining deep-sub-angstrom lateral resolution with depth-sectioning ability, EFMEP provides 3D mapping of hydrogen distribution in metal hydride. These advances offer quantitative insights into hydrogen-metal interactions at the atomic scale and expand the scope of hydrogen characterization in chemically complex materials.

**Benchmarking Hydrogen Detection Limits**

Many metallic elements, including Ti, Zr and Nb, can form stable hydrides (e.g., $TiH_2$, $\delta$-$ZrH_{1.55}$, $NbH_2$) due to their favorable hydrogen absorption thermodynamics (*14, 15, 37*). These hydrides typically adopt a face-centered cubic (FCC) structure with hydrogen occupying tetrahedral interstitial sites, as confirmed by diffraction studies (*14, 15*). To benchmark the hydrogen detection limits of conventional and emerging microscopy techniques, we performed multislice simulations on model FCC $NbH_x$ system (15 nm thickness along the [100] zone axis) with progressively decreasing hydrogen concentrations using a practically applicable high dose of $5 \times 10^5$ e/Å$^2$ (Fig. 1A). At full hydrogen occupancy with 2 hydrogen atoms per metal atom (2.0 H/M), our simulations demonstrate that while ABF, integrated center-of-mass (iCOM, a more quantitative variant of integrated differential phase contrast, iDPC) (*26*), and MEP can all show contrast for hydrogen columns, MEP achieves superior spatial resolution by inversely solving the multiple scattering problem. As hydrogen concentration decreases to 0.54 H/M, the hydrogen columns cannot be observed in both ABF and iCOM images, whereas MEP maintains discernible contrast. Most notably, even at a reduced hydrogen content of 0.36 H/M (corresponding to 12 hydrogen atoms along the beam direction within the simulated column), MEP continues to exhibit observable albeit weakened hydrogen signal, establishing its superior detection limit for low hydrogen concentrations compared to conventional imaging techniques. It should be noted that the hydrogen atoms in the model system used for Fig. 1 are distributed discretely along the z axis. If H atoms form closely packed clusters, the contrast from only two H atoms is detectable with well localized contrast using MEP (Fig. S1).

Fig. 1B further shows the relative intensity of hydrogen and metal columns corresponding to the hydrogen content obtained from different imaging techniques. iCOM images exhibit a linear response at high hydrogen concentrations under the relative thin sample, but with significantly larger uncertainties, especially when the hydrogen content is low. In contrast, MEP shows a consistent linear response with much higher precision, which can also be used to determine the hydrogen content. For ABF, quantitative relationship between the hydrogen contrast and hydrogen content cannot be established, because the contrast is too delocalized and with a low signal-to-noise ratio under this practical dose level. Additionally, iCOM and ABF usually have complicated non-monoclinic contrast and are hindered for quantitative analysis when the sample becomes thicker due to the multiple electron scattering and complicated contrast transfer (*20, 26*).

One of the most crucial experimental parameters for hydrogen imaging in TEM is the dose, because hydrogen atoms are usually highly sensitive to the high energy electron beam due to their high mobility. We compared the dose efficiency of ABF, iCOM, and MEP for hydrogen detection through systematic simulations of a 15 nm-thick $NbH_2$ sample (Fig. S2). Our results demonstrate MEP's superior performance, achieving hydrogen identification at dose levels more than one hundred times lower than required by ABF imaging. At a representative low dose of $5 \times 10^3$ e/Å$^2$, MEP maintains both better resolution and clearer hydrogen signal compared to iCOM. The simulations further reveal that ABF imaging can only detect hydrogen atoms under restrictive conditions requiring both high hydrogen content and elevated dose levels (Fig. S3). These findings unequivocally establish MEP's enhanced sensitivity for hydrogen identification and its significant advantage in dose efficiency over conventional ABF and iCOM approaches.

To evaluate the influence of host metal atomic number on hydrogen detectability, we performed comparative simulations of $TiH_x$ and $NbH_x$ systems. Remarkably, in the $TiH_x$ system, hydrogen columns remain detectable even at ultralow concentrations of 0.18 H/M (corresponding to 6 hydrogen atoms along the beam direction within the simulated column, Fig. S4), demonstrating enhanced sensitivity compared to $NbH_x$. Quantitative analysis reveals a systematically higher relative phase intensity between hydrogen and metal columns in TiH$_x$ (Fig. S5), correlating with improved hydrogen identification capability. This atomic-number-dependent sensitivity suggests that the tail of the host metal's scattering potential has moderate influence on hydrogen detectability in ptychographic imaging.

**Hydrogen Occupancy Sites in TiH$_x$**

We firstly performed experiments in a simple metal hydride, TiH$_x$ using EFMEP. This EFMEP, as a new extension of MEP, filters out most of the signal originating from inelastically scattered electrons (mainly plasmons) and improves convergence of the MEP reconstruction, leading to enhanced sharpness in the contrast of both metal and hydrogen atoms (Fig. S6). Experimental [100] phase images (Fig. 2A) and line profile (Fig. 2C) reveal pronounced contrast at tetrahedral sites. Quantitative analysis yields a relative phase intensity of $(7.4 \pm 1.9)\%$, corresponding to a hydrogen content of $(1.77 \pm 0.45)$ H/M determined by comparing with simulations in similar imaging conditions. We also successfully resolved H-Ti-H triple projections along the [110] zone axis (Fig. 2B and 2D), despite the small separation between Ti and H atomic columns. Importantly, images from both zone-axes consistently agree with the structural model that H atoms exclusively occupy the tetrahedral sites but not the octahedral sites. This model is also consistent with previous reports in TiH$_2$ (*14*).

We further verified the hydrogen occupancy model suggested from the experimental observations in TiH$_x$ via MEP simulations along both [100] and [110] zone axis. The simulation results are presented in Figures 2E-H. The [100] zone axis simulation (Fig. 2E) unambiguously resolves hydrogen columns within the FCC titanium lattice framework, confirming the tetrahedral sites contrast observed experimentally (Fig. 2A). Along the [110] orientation (Fig. 2F), hydrogen columns remain distinguishable despite reduced intercolumn distances, consistent with the resolved H-Ti-H triple projections in experiments (Fig. 2B, 2D). Phase intensity analysis, obtained by averaging across the (200) atomic planes containing Ti-H-Ti columns, reveals distinct peaks between Ti columns (Fig. 2G). The characteristic shoulder features in the [110] line profile (Fig. 2H) further corroborate the hydrogen positions. These simulation results thus validate the experimental findings of exclusive tetrahedral occupancy.

**Resolving Hydrogen Atoms in MPEA hydride**

To further validate the general applicability of MEP, we acquired 4D-STEM data for a multi-principal-element alloy hydride (TiZrHfMoNbH$_x$), a system exhibiting greater complexity than pure metals. MPEAs, defined by near-equiatomic compositions of at least three principal elements, exhibit unique features such as severe lattice distortion, chemical inhomogeneity and short-range order (*27, 38, 39*). These characteristics significantly influence their properties, including hydrogen storage performance (*39-41*). For instance, the TiZrVNbHf MPEA demonstrates a remarkable hydrogen storage capacity of 2.5 H/M at 53 bar and 299 °C, surpassing conventional alloys (*39*). This exceptional capacity suggests hydrogen occupies non-traditional interstitial sites, likely facilitated by the distorted lattice configurations arising from atomic size mismatch. Neutron diffraction studies have confirmed the occupation of octahedral and tetrahedral interstitial sites simultaneously in this MPEA (*40*). The behavior of hydrogen within MPEA hydrides is further complicated by the lattice distortions and inherent chemical inhomogeneities (*29, 40, 41*).

The TiZrHfMoNb MPEA used in this study (chemical composition in Table S1) crystallizes in a BCC structure and, after activation at 250 °C, exhibits fast hydrogen absorption kinetics and a high hydrogen storage capacity of ~1.6 H/M (*42*). We first verified the MPEA and its hydride by using electron energy loss spectroscopy (EELS). The low-energy-loss part revealed plasmon peaks at 19.3 eV for TiZrHfMoNb and 21.0 eV for TiZrHfMoNbH$_x$ (Fig. S7). Such plasmon peak shifts between alloys and their hydrides are well-documented in metallic systems (*43-45*). For instance, prior studies reported peaks at 17.6 eV for Ti and 19.4 eV for γ-TiH (*23*). This consistent shift confirms the successful formation of MPEA hydrides. The superior spatial resolution and hydrogen detection sensitivity of MEP reconstruction are also demonstrated by comparing to conventional techniques (HAADF, ABF, and iCOM in Fig. S8). Figure 3 displays two distinct regions along the [100] zone axis, both showing clear interstitial contrast (Fig. 3A, B), though with varying phase intensities. The weaker contrast in Region 1 (Fig. 3A) compared to Region 2 (Fig. 3B) suggests regional hydrogen content differences, likely arising from MPEA chemical inhomogeneity. Line profile analysis revealed peaks corresponding to hydrogen columns between metal atoms (Fig. 3C, D), with a stronger peak in Fig. 3D. Statistical evaluation indicated that the phase intensity of hydrogen columns was ~3.0% and ~4.8% of metal columns (Fig. 3E, F), respectively. Based on the mean atomic number (43.08, close to Nb, Z = 41), the estimated hydrogen contents were ~0.88 H/M and ~1.38 H/M. However, the more pronounced positional disorder of hydrogen atoms in the MPEA hydride compared to metal atoms inherently causes an underestimate of hydrogen content in the estimation using projected phase intensity.

By comparing the atomic structure models of BCC and FCC lattices showing tetrahedral and octahedral interstitial sites (Fig. S9), we find the hydrogen occupies exclusively at tetrahedral interstitial sites, with no detectable occupancy at octahedral sites, which is further verified via the MEP reconstruction along [110] zone axis (Fig. S10). This finding is in agreement with previous neutron powder diffraction (NPD) studies, confirming the preferential tetrahedral site occupancy as a fundamental characteristic of hydrogen incorporation in this system (*46*). The consistency between real-space atomic-resolution MEP results and bulk-averaged NPD data provides strong validation for both techniques while highlighting the unique capability of MEP to directly visualize hydrogen positions at the atomic scale.

MEP achieves sub-picometer precision in atomic positioning of heavy metal elements, as previously demonstrated (*31*), enabling precise lattice distortion measurements in MPEA hydrides. Figures 3G and H display statistical analyses of nearest M-M and H-M distances in the two regions. The in-plane projected M-M distances, measured from the projected phase images, are 231.6 ± 1.8 pm and 235.0 ± 2.3 pm for the two regions, respectively. The difference in the average M-M distances between the two regions primarily reflects their differing hydrogen contents. Similarly, the in-plane projected H-M distances were 164.1 ± 7.0 pm and 166.1 ± 6.5 pm, respectively. Based on the geometry of hydrogen occupying tetrahedral interstitial sites in an FCC lattice viewed along the [100] zone axis, the corresponding calculated three-dimensional H-M distances are 200.9 ± 8.6 pm and 203.4 ± 8.0 pm, respectively. Notably, the observed standard deviations reflect both methodological uncertainties and intrinsic lattice distortions. The actual positional precision should be better than these deviations. The larger standard deviation in H-M distances arises from the pronounced displacement of hydrogen atoms, which is more significant than that of metal atoms. Crucially, MEP achieves reliable picometer-level precision for hydrogen positions, which is unattainable with conventional techniques (*47*). Since the projected phase image represents an average of all depth slices, these measurements inherently smooth out local atomic displacements. By analyzing each slice (Fig. S11), we obtained M-M distances of 231.7 ± 4.2 pm and 234.9 ± 3.8 pm, corresponding to 1.8% and 1.6% variation, respectively. This approach not only mitigates the smearing effect of projection but also provides direct evidence of lattice distortions. Furthermore, our results also demonstrate that hydrogen columns remain clearly detectable even in thicker specimens of approximately 30 nm (Fig. S12), where we achieved an upper limit of precision of 7.0 pm for H-M distances.

For comparative analysis, we performed MEP reconstruction for the pristine MPEA (Fig. S13). The phase image along the [100] zone axis reveals no detectable contrast at interstitial sites (Fig. S13A), consistent with the absence of hydrogen occupation. Corresponding line profiles further confirmed the lack of peaks between metal columns (Fig. S13B). The nearest in-plane projected M-M distance was determined to be 237.2 ± 2.5 pm from the projected phase image (Fig. S13C). Based on the BCC crystal structure and the projected atomic positions, the calculated three-dimensional distances from metal atom positions to the centers of octahedral and tetrahedral interstitial sites are approximately 167.7 ± 1.8 pm and 187.5 ± 2.0 pm, respectively. While hydrogen absorption typically induces lattice expansion within a given crystal structure, the FCC structure adopted by the hydride phase inherently possesses a different atomic packing density compared to the BCC structure of the pristine alloy. The tetrahedral interstitial sites in FCC lattices are significantly larger than both octahedral and tetrahedral sites in BCC lattices, facilitating the accommodation of hydrogen atoms and contributes to the distinct atomic configurations and measured distances observed between the two phases. The smaller uncertainty (~2 pm) compared to the FCC H-M distances (8.6 pm) also reflects the enhanced disorder of H position in MPEA hydrides.

**Three-Dimensional Inhomogeneities of Hydrogen Atoms**

Depth-resolved analysis through MEP reconstruction provided additional insights into the three-dimensional structural characteristics and hydrogen distribution. Phase images from individual depth slices distinctly revealed both lattice distortions and chemical inhomogeneities of metal sites in three dimensions (Fig. S13D and S13E). The slice-by-slice measurements yield an average M-M distance of 237.3 ± 6.1 pm, representing a 2.6% variation in pristine phase (Fig. S13F), which is greater than that observed in the hydride phase (1.8% or 1.6%). These comparative results suggest that hydrogen incorporation within the FCC hydride phase reduces the intrinsic lattice distortion present in the pristine BCC MPEA. The relief of lattice distortion by deuterium incorporation has also been observed in TiVNb based MPEAs using total scattering measurements (X-ray and neutron diffraction) and Reverse Monte Carlo structure modelling (*40*). This distortion relief may arise from three interconnected factors: hydrogen attracts strong hydride-forming elements (e.g., Ti, Zr, Hf), driving localized atomic rearrangement that reduces chemical disorder; the BCC to FCC phase transition partially resets pre-existing lattice distortions; and hydrogen's relatively uniform tetrahedral site occupancy helps distribute strain evenly. Collectively, these effects counterintuitively alleviate—rather than exacerbate—lattice distortion in MPEAs.

The large lattice distortions and chemical inhomogeneities in the MPEA matrix can necessarily lead to non-uniform hydrogen distribution in the hydride phase. To improve the signal-to-noise level of H sites, the depth dependent structural analysis was performed by summing the phase from 5 slices from MEP reconstruction (1 nm thickness per slice) with three representative depth ranges shown in Figure 4A. Phase contrast variations across different depth regions demonstrate the significant heterogeneous distribution of hydrogen atoms at tetrahedral interstitial sites. The depth profile (Fig. 4B) illustrates both the inhomogeneous distribution and the displacement of metal atoms. Notably, the magnitude of this metal atom displacement appears significantly lower than that observed in the pristine MPEA matrix. Furthermore, the phase intensity fluctuations of hydrogen columns along the depth direction reveal substantial three-dimensional inhomogeneity in hydrogen filling. Hydrogen atoms exhibit significantly greater displacements compared to metal atoms, consistent with their lower atomic mass and weaker lattice constraints. These findings provide direct experimental evidence for three-dimensionally heterogeneous hydrogen distribution in MPEA hydrides, arising from the combined effects of intrinsic lattice distortions and chemical inhomogeneities in the host alloy matrix.

The inhomogeneity of hydrogen concentration variations along the depth direction reveals is further quantified using the peak intensity of H sites, as demonstrated for three characteristic depth ranges in Figure 4C. While hydrogen columns are detectable at all depths, their phase intensity shows remarkable fluctuations (Fig. 4D), indicating partial and variable occupancy of interstitial sites. However, due to the similar atomic numbers of Mo, Nb, and Zr, MEP cannot reliably distinguish these elements in our analysis. This limitation prevents direct correlation between local hydrogen occupancy and specific elemental environments. Specifically, the 10-14 nm depth region exhibits 30-44% higher relative phase intensity compared to the 5-9 nm and 15-19 nm regions. Statistical analysis of relative phase intensities shows hydrogen column intensities of 5.0%, 6.5%, and 4.5% of metal column intensities for these respective depth ranges (Fig. 4E-G). These results suggest that hydrogen near the surface is more prone to release due to reduced coordination constraints.

**Conclusions**

In this study, we have successfully demonstrated the capability of multislice electron ptychography to directly detect hydrogen atoms at atomic resolution, as evidenced by our investigations of two distinct hydride systems. Through the adaption of energy-filtered multislice electron ptychography, we achieved quantitative determination of hydrogen content in both materials. Our depth-resolved structural analysis reveals the three-dimensional inhomogeneous distribution and the displacement of hydrogen atoms. The significant displacement of hydrogen atoms is further corroborated by precise measurements of H-M bond lengths with picometer-level precision, which surpasses the capabilities of conventional characterization techniques. These results highlight MEP's exceptional potential as a powerful imaging tool for materials research, particularly in hydride systems with local lattice distortion and inhomogeneity. Our work establishes a foundation for deeper investigations of hydrogen behavior in solids, enabling future studies to extract critical information about local site occupancy, atomic vibrations, and hydrogen mobility from MEP data. Importantly, the methodology developed here is not limited to hydride systems, but can be readily extended to a broad range of materials containing light elements adjacent to heavy ones, including but not limited to oxides, nitrides, carbides, and borides. This versatility positions MEP as a transformative technique for atomic-scale characterization in materials science.

**Acknowledgments:** We thank Tieqiao Zhang and Chang Lu for assistance during the experiments. This work used the facilities of the National Center for Electron Microscopy in Beijing at Tsinghua University. The authors acknowledge funding supports from the National Key Research and Development Program of China (MOST, Grant Nos. 2023YFA1406400 and 2022YFA1405100), the National Natural Science Foundation of China (Grant Nos. 52273227 and U22A6005), and Guangdong Major Project of Basic Research, China (Grant No. 2021B0301030003).

**Funding:**

National Key Research and Development Program of China (MOST, Grant Nos. 2023YFA1406400 and 2022YFA1405100)

National Natural Science Foundation of China (Grant Nos. 52273227 and U22A6005)

Guangdong Major Project of Basic Research, China (Grant No. 2021B0301030003)

**Author contributions:** Conceptualization: PCL, ZC; Methodology: PCL, CLP, ZHD, ZXS, TL, CC, HHS, ZC; Investigation: PCL, CLP; Visualization: PCL; Supervision: ZC, LG; Writing – original draft: PCL; Writing – review & editing: CZ.

**Competing interests:** Authors declare that they have no competing interests.

**Data and materials availability:** Raw experimental data of the results presented in the main text will be available after the manuscript is accepted. The code for multislice electron ptychography is adapted from the previously published versions in Zenodo at https://doi.org/10.5281/zenodo.4659690 (*48*).

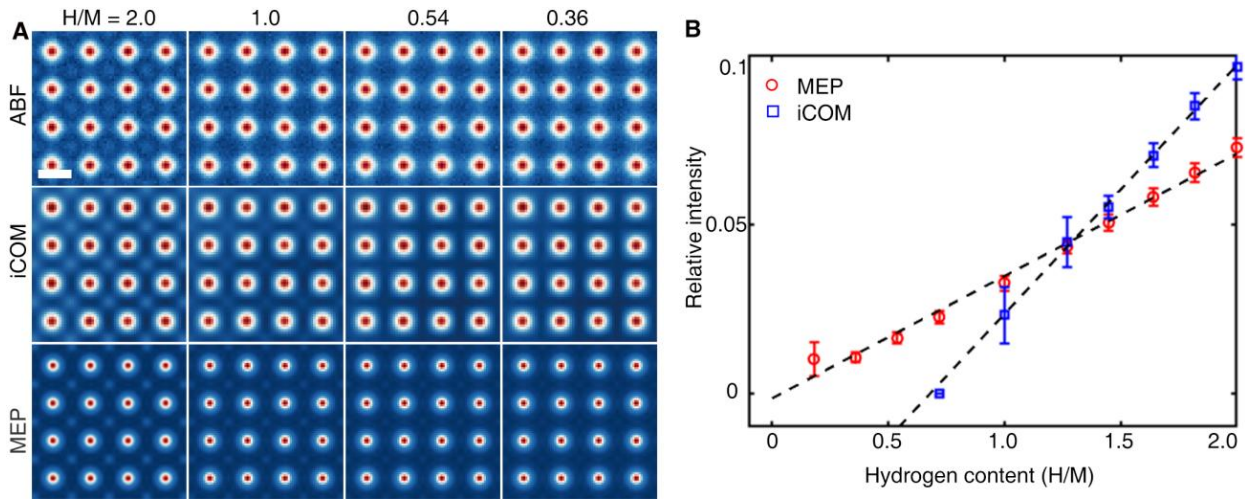

**Fig. 1. Hydrogen detection sensitivity across different imaging techniques.** (**A**) Simulated images of NbH$_x$ (15 nm thickness) with varying hydrogen content over Nb metal atoms (0.36-2.0 H/M) acquired by contrast-inverted annular bright-field (ABF), integrated center-of-mass (iCOM), and multislice electron ptychography (MEP). Scale bar, 2 Å. (**B**) Hydrogen-to-metal column intensities vs hydrogen content for iCOM and MEP, demonstrating MEP's superior detection sensitivity.

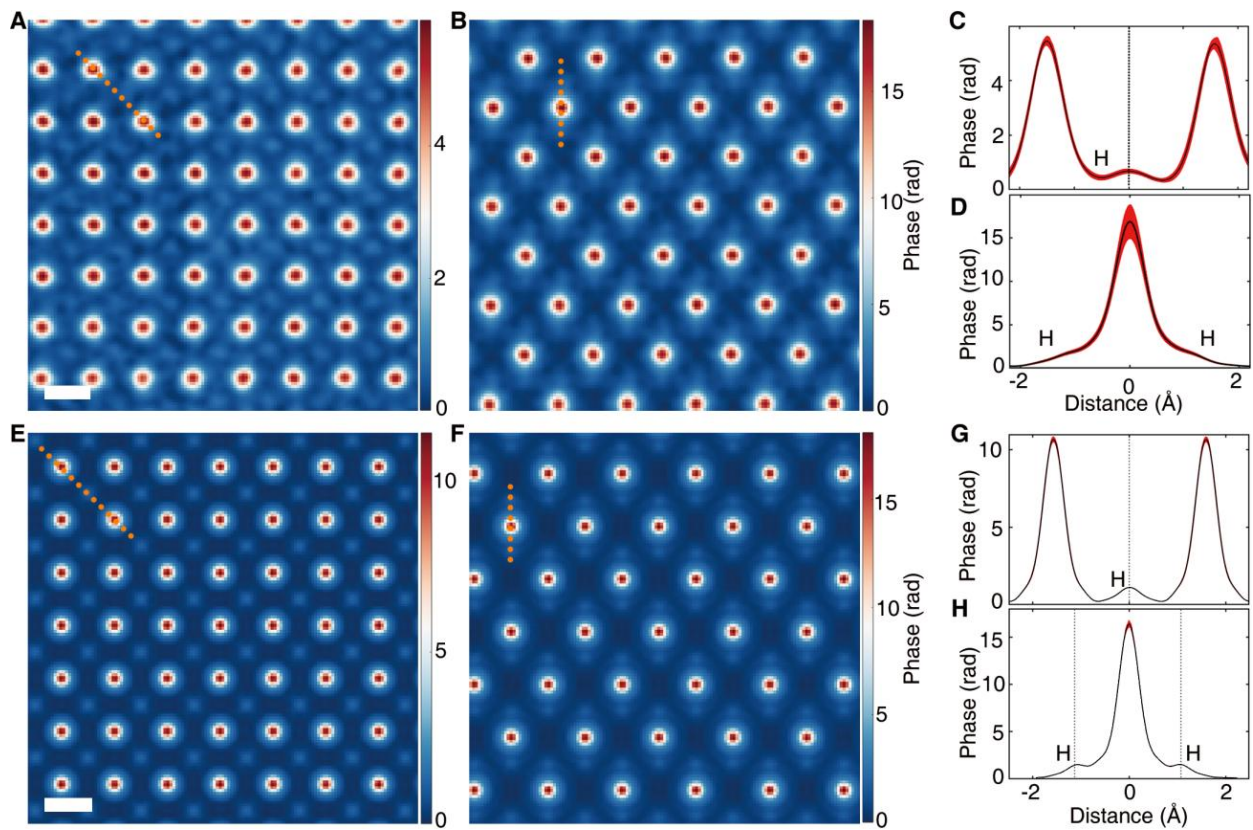

**Fig. 2. Atomic-resolution imaging of TiH$_2$ by multislice electron ptychography.** (**A** and **B**) Phase images along [100] and [110] zone axis using experimental data, summing only the middle slices with a thickness of 11 nm to 22 nm, respectively to exclude surface damaged slices. (**C** and **D**) Experimental phase profiles across Ti-H-Ti and H-Ti-H columns, from positions indicated by orange dashed lines in (A) and (B), respectively. (**E** and **F**) Total phase images reconstructed from simulated data along [100] and [110] zone axis, respectively, obtained by summing all depth slices. (**G** and **H**) Corresponding phase profiles across Ti-H-Ti and H-Ti-H columns, with red shading indicating variation from ~30 individual profiles and black lines showing their averages. Profile locations are marked by orange dashed lines in (E) and (F). Scale bars in (A) and (E), 2 Å.

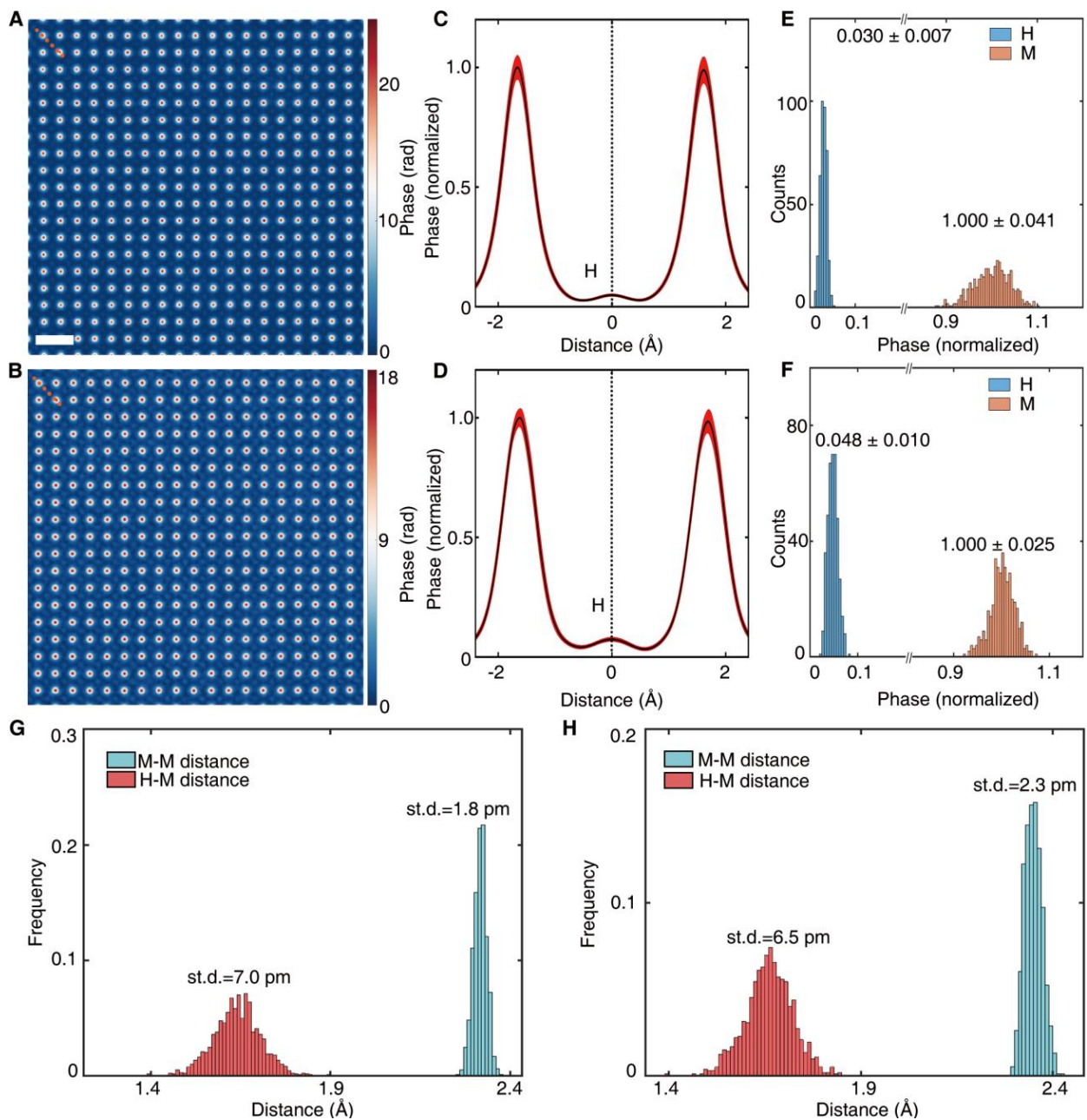

**Fig. 3. Imaging hydrogen atoms in TiZrHfMoNbH$_x$ by multislice electron ptychography.** (**A** and **B**) Phase images from two distinct regions summed using middle slices (23 nm and 20 nm thickness, respectively) to eliminate surface artifacts. Scale bar, 5 Å. (**C** and **D**) Phase profiles across M-H-M columns, with red shading indicating variations from ~100 individual profiles and black lines showing their averages. Measurement positions of one example are marked by orange dashed lines in (A and B). (**E** and **F**) Phase value distributions for hydrogen and metal sites in corresponding regions. (**G** and **H**) Histograms of projected M-M and H-M atomic distance. st.d., standard deviation.

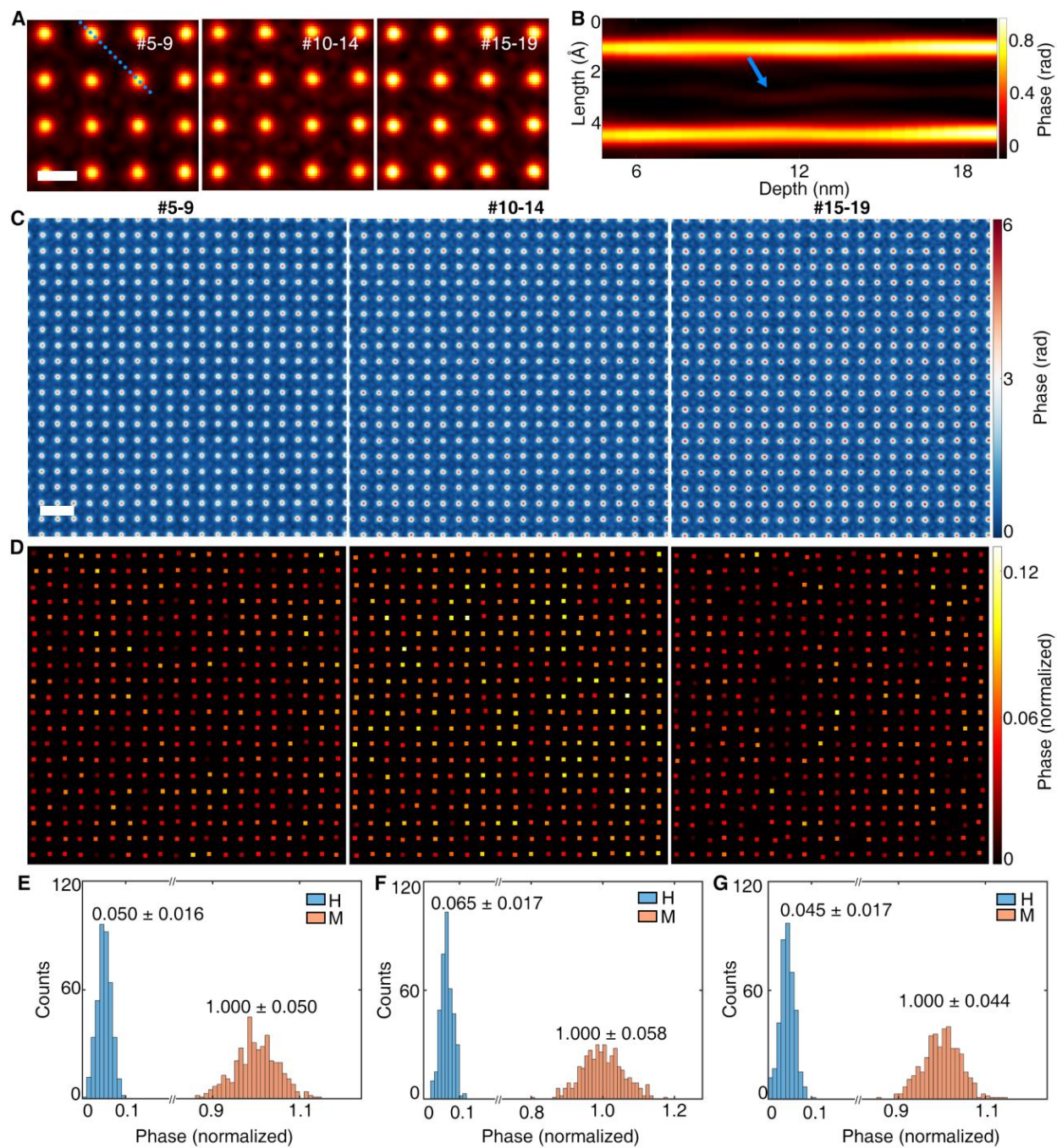

**Fig. 4. Depth-sectioning analysis of hydrogen distribution in TiZrHfMoNbH$_x$ via multislice electron ptychography.** (**A**) Phase images from three distinct depth regions, each integrated from 5 consecutive slices (5-9 nm, 10-14 nm, and 15-19 nm depth ranges). (**B**) Depth profile of phase intensity along the M-H-M direction indicated by the blue dashed line in (A). An arrow denotes the displacement of hydrogen. (**C**) Expanded field-of-view phase images of the same three depth regions shown in (A). (**D**) Corresponding hydrogen column phase intensity maps for each integrated depth region shown in (C). (**E-G**) Phase value distributions comparing hydrogen and metal sites across the three depth regions. Scale bars, 2 Å (A) and 5 Å (C).

# Supplementary Materials

**Materials and Methods**

Sample preparation

The TiZrHfMoNb MPEA was prepared by arc melting under an Ar atmosphere (*42*). The alloy ingot was remelted six times to improve its homogeneity. Each specimen was machined into square plates (10 × 10 × 1 mm) using wire electrical discharge machining, ground with silicon carbide (SiC) abrasive paper, and polished using a vibratory polisher.

TEM lamellas were extracted from the TiZrHfMoNb MPEA plate using $Ga^+$ ion beam via the standard lift-out method in a focused ion beam (FIB) system (Zeiss Auriga). Subsequently, the lamella was hydrogenated at 250°C under 1 bar hydrogen pressure for 30 minutes. To mitigate FIB-induced damage caused by $Ga^+$ ion bombardment and prevent hydrogen release due to FIB milling heating effects, a flash electrochemical polishing technique was applied (*49*). The polishing was conducted at −30 °C in a solution of 4% perchloric acid and 96% alcohol at 14 V for 0.1 s.

Multislice simulations

ABF, iCOM, and MEP are STEM techniques capable of resolving light elements. To evaluate their performance in hydrogen imaging, we simulated ABF, iCOM, and MEP images using an $NbH_x$ model system and the multislice simulation package abTEM (*50*). Experimental parameters, including beam energy and probe-forming semi-angle, were applied in the simulations. The sample thickness was set to 15 nm, and hydrogen content was systematically varied by uniformly removing hydrogen atoms along the *z*-direction from $NbH_2$. The Debye-Waller factors used were 0.43 $Å^2$ for Nb (*51*) and 1.66 $Å^2$ for H (*52*). For MEP simulations, the probe was focused 20 nm above the sample, and 60 × 60 diffraction patterns were generated with a step size of 0.5 Å × 0.5 Å. Each pattern had 512 × 512 pixels at 0.616 mrad/pixel, which were later cropped to 256 × 256 pixels during reconstruction to match the experimental $k_{max}$ values. For iCOM, 200 × 200 diffraction patterns were simulated, each with 256 × 256 pixels at 0.616 mrad/pixel. In iCOM, a defocus value of −3 nm and a collection angle of 0-24.3 mrad were applied. For ABF, the probe was focused on the top of the sample to obtain the best contrast, with a collection angle of 12.15-24.3 mrad. Poisson noise was added to model the finite illumination dose conditions. For $TiH_x$, the Debye-Waller factors were 0.52 $Å^2$ for Ti (*51*), while hydrogen retained the same values as in $NbH_x$. Unless otherwise specified, the illumination dose was fixed at $5 \times 10^5$ e/$Å^2$.

4D-STEM experimental data acquisition

TEM experiments were conducted using an aberration-corrected JEM-ARM300F2 transmission electron microscope operated at 300 keV with a probe-forming aperture of 24.3 mrad. The 4D-STEM datasets were acquired using a Gatan K3 camera coupled with the STEMx modular system. Unless otherwise specified, the experiments employed a defocused electron probe positioned approximately 20 nm above the sample's top surface. Each 4D-STEM dataset consisted of 100 × 100 uniformly distributed scanning points with a step size of 0.52 Å, using a beam current of approximately 10 pA. The original 512 × 512-pixel diffraction patterns were binned down to 128 × 128 pixels prior to ptychographic reconstruction. To enhance real-space sampling, the diffraction patterns were padded to 256 pixels, achieving a maximum spatial frequency ($k_{max}$) of 7.967 $Å^{-1}$.

Reconstructions were performed with a slice thickness of 1 nm. For the thicker TiZrHfMoNb hydride region (30 nm thickness), the probe was overfocused by about 10 nm above the sample surface, and the scan step size was reduced to 0.42 Å.

It should be noted that neither the MEP algorithms nor the simulations accounted for inelastic electron scattering (*34*). Typically, inelastically scattered electrons contribute incoherently to diffraction patterns, leading to image blurring and degraded reconstruction quality. To address this issue, we employed an energy slit with a 5-eV full width centered at zero-loss energy to mitigate plasmonic scattering contributions from the diffraction patterns. As demonstrated in Fig. S6, this adjustment revealed finer structural features in the diffraction patterns. By filtering out approximately 30% of the total signal originating from inelastically scattered electrons, we improved the convergence of the MEP reconstruction. Consequently, for samples with approximately 18 nm thickness and substantial surface amorphous layers, both metal and hydrogen atoms exhibited significantly enhanced sharpness in the filtered MEP reconstruction compared to unfiltered results.

Electron energy loss spectroscopy (EELS)

We performed electron energy loss spectroscopy (EELS) characterization on both the pristine TiZrHfMoNb MPEA and its hydride using an aberration-corrected JEM-ARM300F2 transmission electron microscope operating at 300 kV, equipped with a Gatan K3 camera. The EELS experiments were conducted with a beam current of 10 pA, employing a convergence semi-angle of 24.3 mrad and a collection semi-angle of 130 mrad, while maintaining an energy dispersion of 0.09 eV per channel.

Statistical analyses

The atoms in the MEP reconstructed images were all fitted by a two-dimensional anisotropic Gaussian function, $f(x) = A \exp\left[-\frac{(x-\mu_x)^2}{2\sigma_x^2} - \frac{(y-\mu_y)^2}{2\sigma_y^2}\right] + B$, where the phase of each atom is proportional to parameter $A$. The relative phase intensity between hydrogen columns and metal columns is $\frac{\varphi_H}{\varphi_M}$, where $\varphi_H$ and $\varphi_H$ is the phase of hydrogen columns and metal columns, respectively. The uncertainty in relative phase intensity is derived according to propagation of errors: $\frac{\varphi_H}{\varphi_M}\sqrt{\left(\frac{\sigma_{\varphi_H}}{\varphi_H}\right)^2 + \left(\frac{\sigma_{\varphi_M}}{\varphi_M}\right)^2}$, where $\sigma_{\varphi_H}$ and $\sigma_{\varphi_M}$ are standard deviations for $\varphi_H$ and $\varphi_H$, respectively. After atom positions were determined by fitting, M-M interatomic distances were measured directly from the projected phase images and within individual slices. However, owing to the weak phase intensity of hydrogen atoms and artifacts of single slice, only H-M distances from the projected phase are used in the statistical analyses.

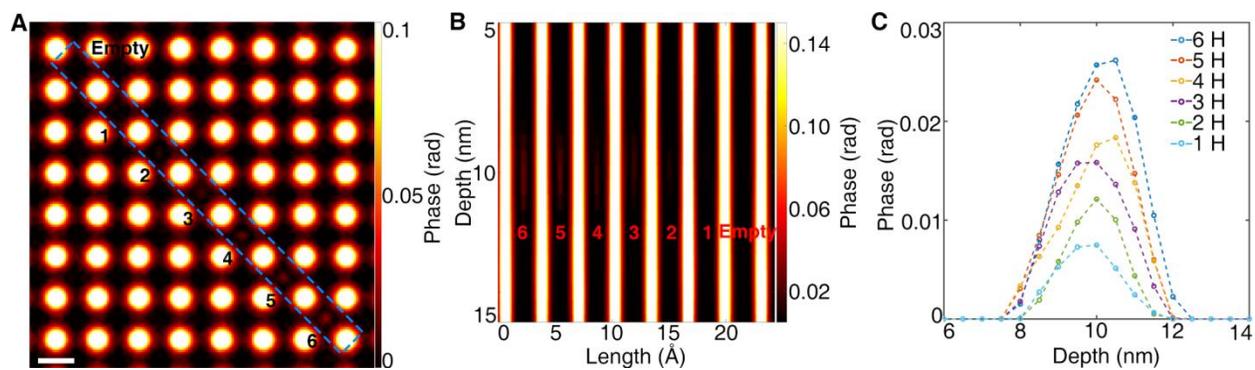

**Fig. S1. Simulated ptychographic reconstruction of hydrogen clusters in FCC Nb.** (**A**) Phase reconstruction of the centroid slice (showing optimal hydrogen contrast), revealing clusters with varying hydrogen atoms. Scale bar, 2 Å. (**B**) Depth profile along the blue dashed rectangle in (A). (**C**) Cluster-size-dependent phase intensity in depth profile.

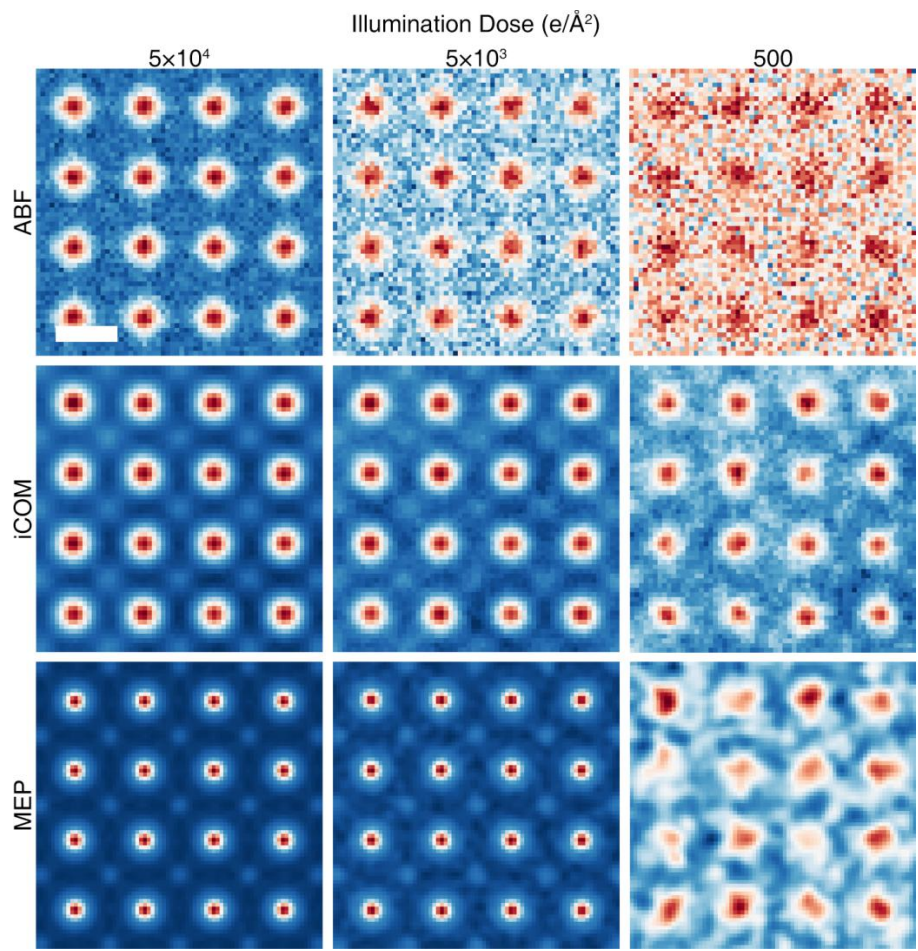

**Fig. S2. Comparative simulations of imaging techniques under low-dose conditions.** Multislice electron ptychography (MEP), annular bright-field (ABF), and integrated center-of-mass (iCOM) simulations for a 15 nm thick $NbH_2$ sample. Scale bar, 2 Å.

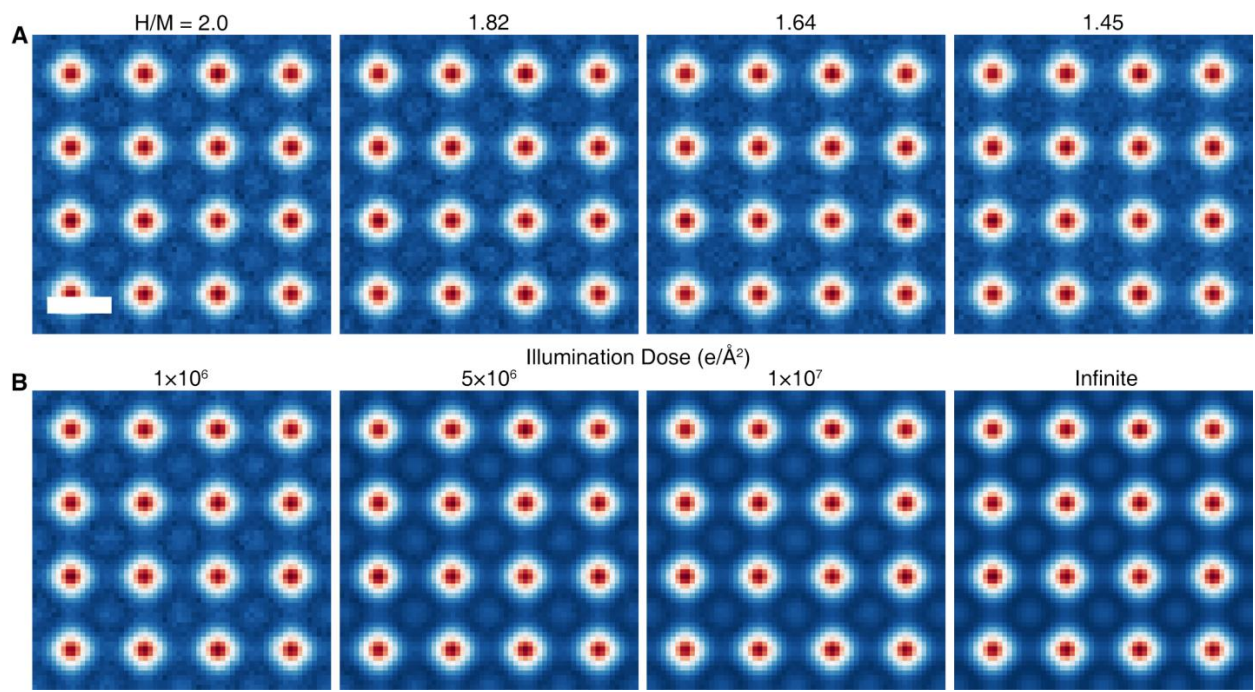

**Fig. S3. Hydrogen content and dose dependence in ABF imaging simulations of NbH$_2$.** (**A**) Simulated ABF images showing hydrogen visibility variation with hydrogen content (0.18-1.0 H/M) at fixed dose (5×10$^5$ e/Å$^2$ ). (**B**) Dose-dependent ABF image series of NbH$_2$. Scale bars, 2 Å.

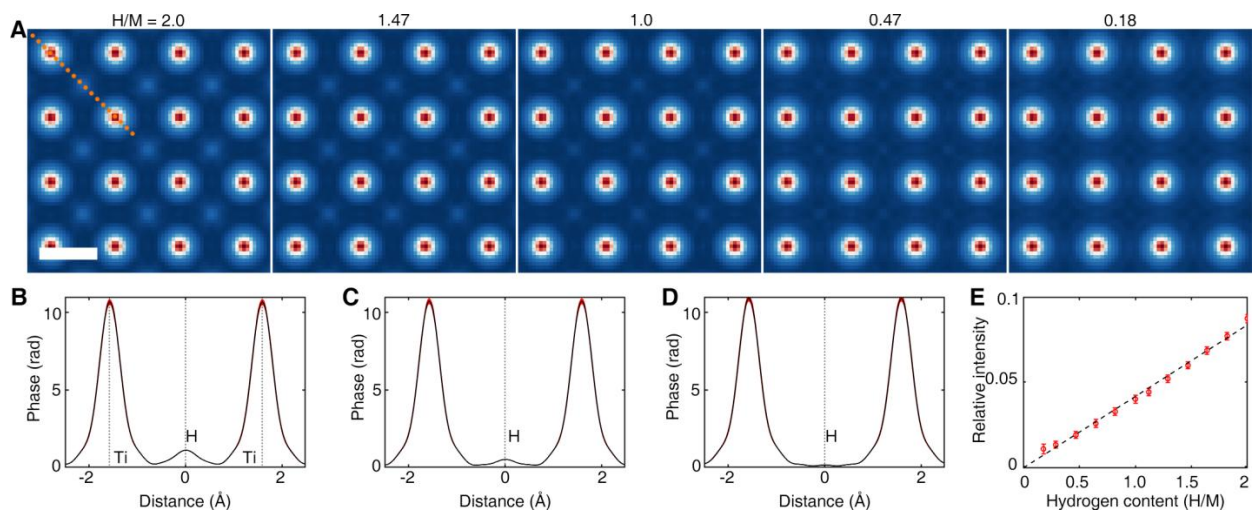

**Fig. S4. Quantitative hydrogen content analysis in TiH$_x$ via multislice electron ptychography simulations.** (**A**) Reconstructed phase images of TiH$_x$ with varying hydrogen content (0.18-2.0 H/M). Scale bar, 2 Å. (**B-D**) Phase profiles across Ti-*H-Ti* columns at hydrogen contents of (B) 2.0, (C) 1.0, and (D) 0.18 H/M. Red shading indicates variations from ~80 individual profiles; black lines represent the averages. Measurement positions marked by orange dashed lines in (A). (**E**) Linear correlation between hydrogen-to-metal phase intensities and hydrogen content, validating MEP's quantitative capability.

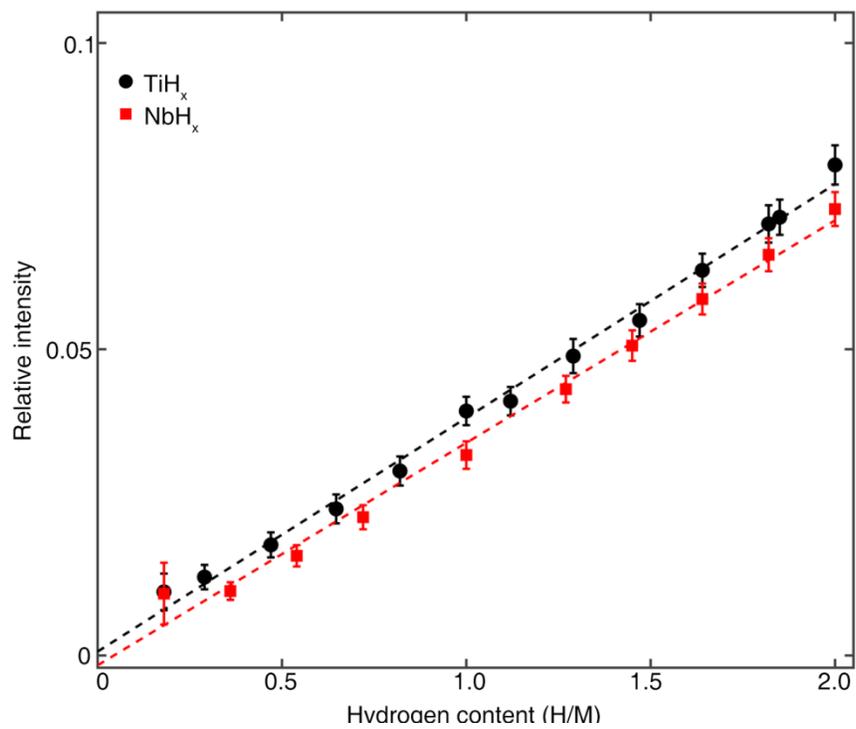

**Fig. S5. Hydrogen content vs relative phase intensity via MEP in TiH$_x$ and NbH$_x$.**

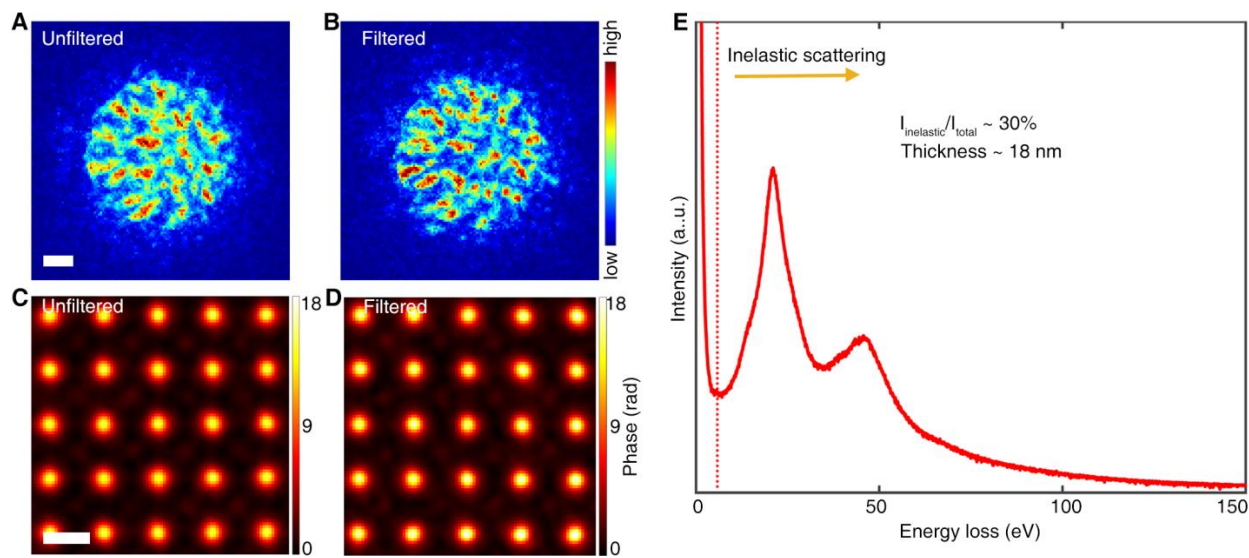

**Fig. S6. Ptychographic reconstruction from energy unfiltered and filtered 4D-STEM data for TiZrHfMoNbH$_x$ (~18 nm thickness).** Representative diffraction patterns from 4D-STEM datasets without (**A**) and with (**B**) energy filtering, showing reduced blurring after filtering. Scale bars, 10 mrad. (**C** and **D**) MEP reconstructions using unfiltered (**C**) and filtered (**D**) data, demonstrating improved hydrogen column sharpness after filtering, enabling more accurate hydrogen quantification. Scale bar, 2 Å. (**E**) Low-loss EELS spectrum from the analyzed region, showing plasmon peaks accounting for ~30% of total scattering. The measured $t/\lambda = 0.41$ indicates ~15 nm amorphous layer in thickness.

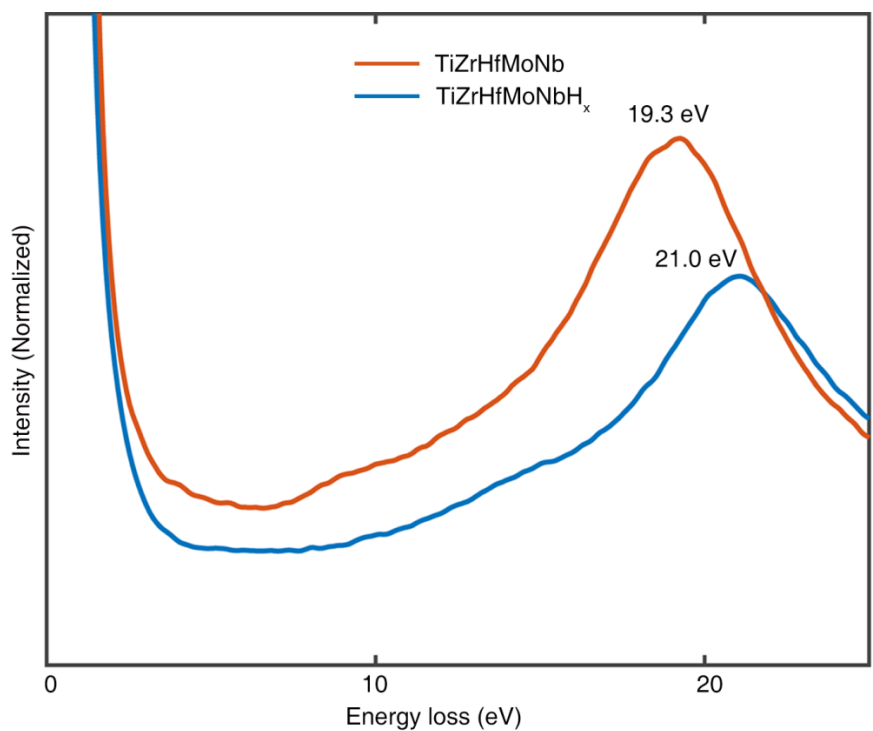

**Fig. S7. Hydrogenation-induced plasmon evolution in TiZrHfMoNb MPEA.** Low-loss EELS spectra showing a 1.7 eV plasmon shift from 19.3 eV (pristine MPEA) to 21.0 eV (hydride phase).

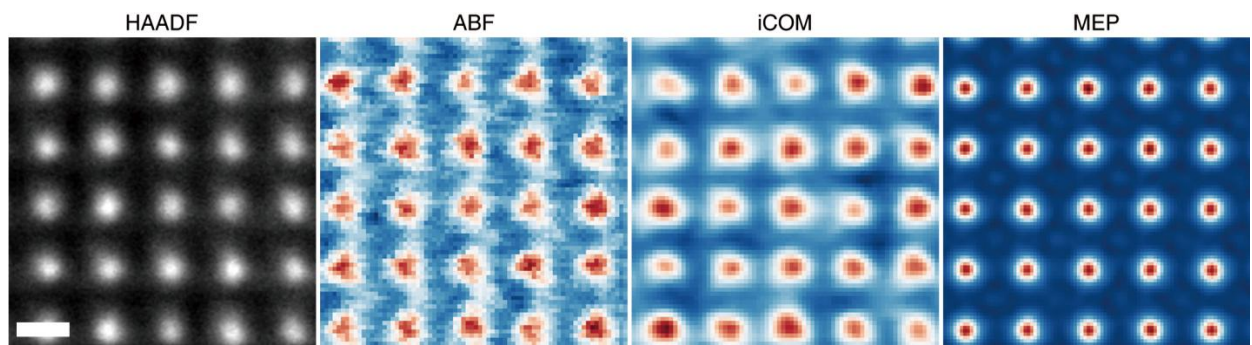

**Fig. S8. Atomic-scale imaging of TiZrHfMoNbH$_x$ with HAADF, ABF, iCOM, and MEP. Scale bar, 2 Å.**

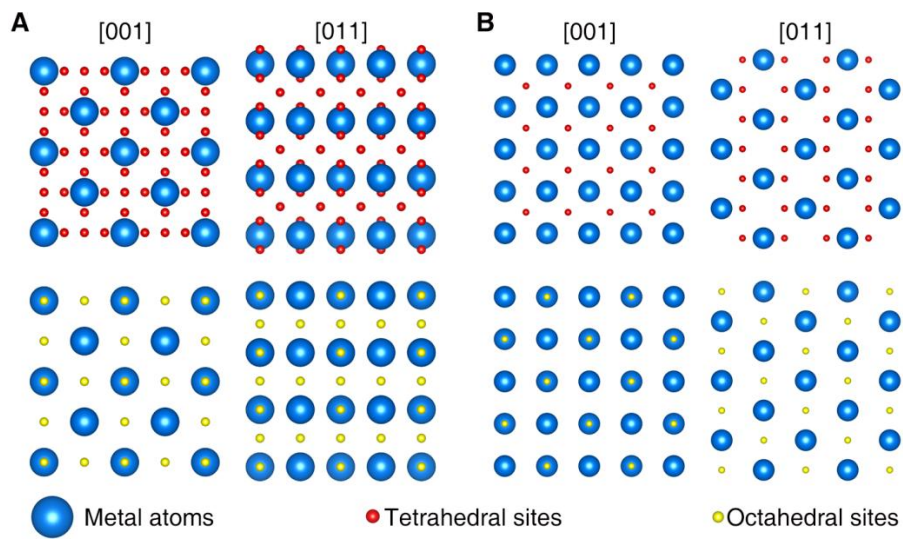

**Fig. S9. Atomic structure models of interstitial sites in BCC and FCC lattices.** (**A**) BCC viewed along the [001] and [011] zone axis. (**B**) BCC viewed along the [001] and [011] zone axis.

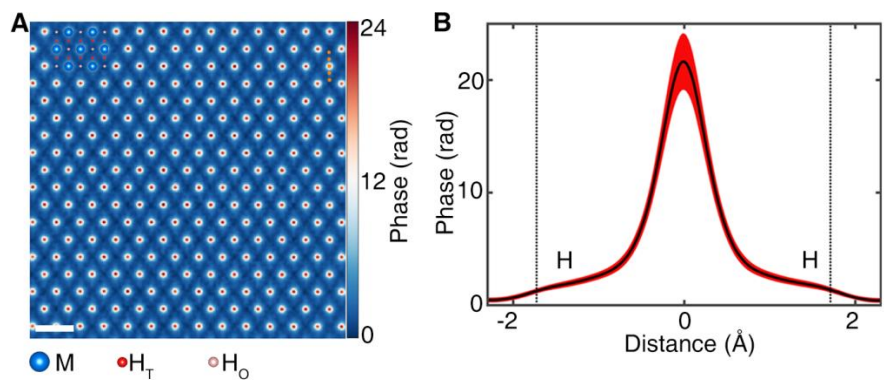

**Fig. S10. Multislice electron ptychographic reconstruction of TiZrHfMoNbH$_x$ along [110] zone axis.** (**A**) Phase image from middle slices (22 nm thick in total) showing hydrogen contrast at tetrahedral sites with undetectable signal at octahedral positions. Scale bar, 2 Å. (**B**) Phase profiles across H-Ti-H triplets.

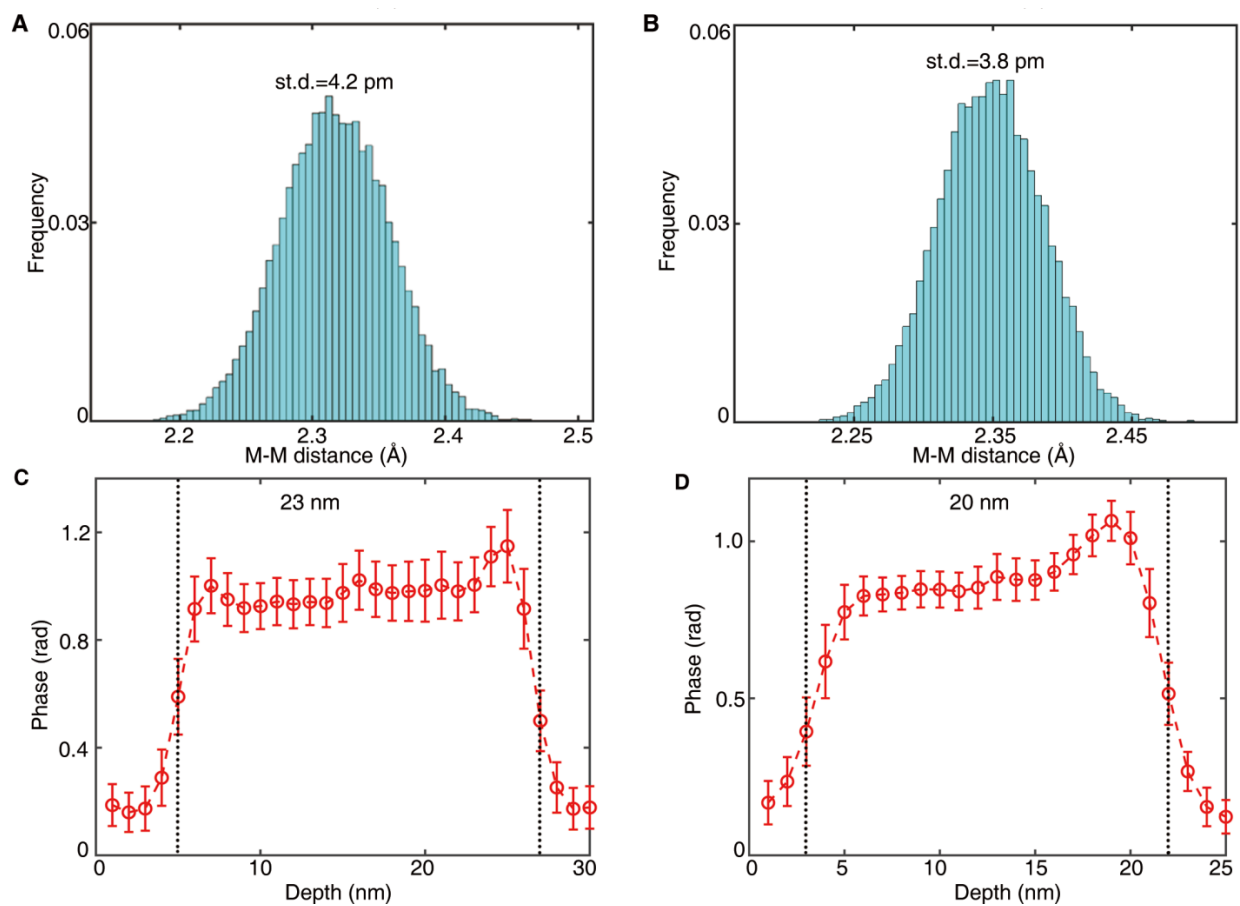

**Fig. S11. Bond length statistics and thickness measurement in TiZrHfMoNbH$_x$ via MEP.** (**A** and **B**) Histograms of M-M bond lengths aggregated from all depth slices in regions corresponding to Fig. 3A and Fig. 3B. (**C** and **D**) Depth phase profiles of metal atom columns with sample thicknesses of ~23 nm (C) and ~20 nm (D), corresponding to regions in Fig. 3A, B. The error bars and the data points are the standard deviation and mean values of the phase intensity.

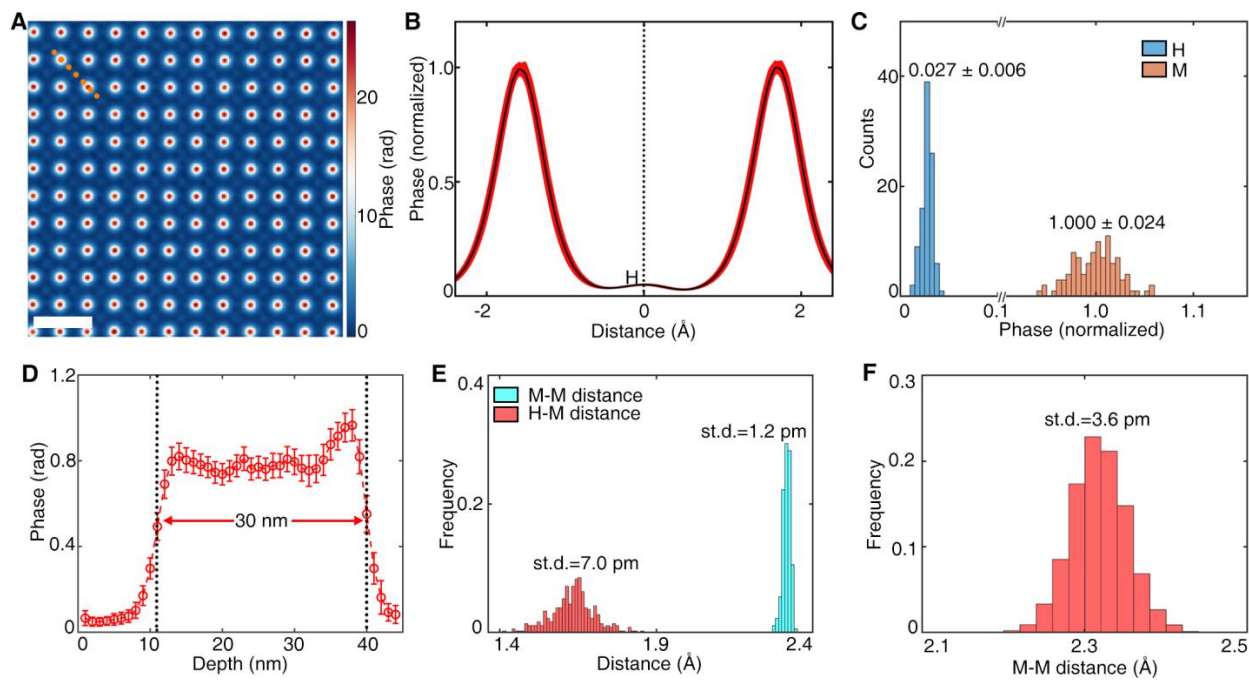

**Fig. S12. Multislice electron ptychographic reconstruction of TiZrHfMoNbH$_x$ from a thicker sample (30 nm thick in total).** (**A**) Phase image reconstructed from middle slices showing both metal and hydrogen columns. Scale bar, 5 Å. (**B**) Phase profiles across M-H-M columns (orange dashed lines in A), with red shading indicating variations from ~80 atomic columns and black lines showing averages. (**C**) Phase value distributions for hydrogen and metal sites. (**D**) Depth phase profile of metal atoms revealing ~30 nm sample thickness. The error bars and the data points are the standard deviation and mean values of the phase intensity. (**E**) Bond length distributions for M-M and H-M. (**F**) Histograms of M-M bond lengths aggregated from all depth slices.

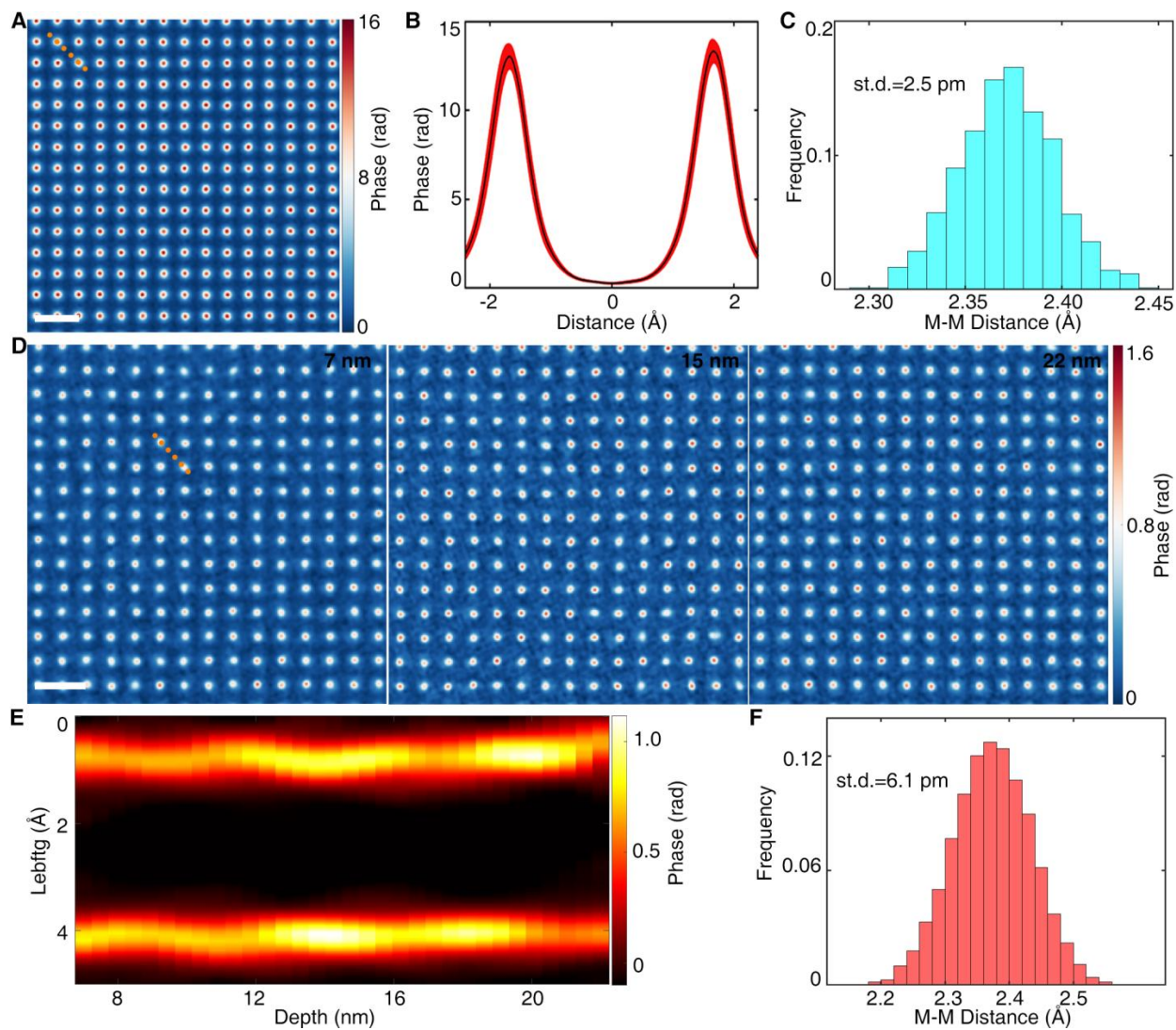

**Fig. S13. Three-dimensional lattice distortion and chemical inhomogeneity analysis in TiZrHfMoNb MPEA via multislice electron ptychography.** (**A**) Phase image from middle slices (14 nm thick in total) to exclude surface artifacts. (**B**) Phase profiles along the orange dashed line in (A), with red shading showing variations from ~100 atomic columns and black lines indicating the averages. (**C**) Histogram of projected M-M bond lengths. (**D**) Depth-resolved phase images from three selected slices, revealing severe lattice distortions and chemical inhomogeneities. (**E**) Depth profile along the orange dashed line in (D), showing z-direction atomic displacements and chemical inhomogeneities. (**F**) Histograms of M-M bond lengths aggregated from all depth slices.

**Table S1. Elemental properties and chemical composition of TiZrHfMoNb (Atomic number Z, atomic radius r, chemical composition in at.% and wt.%).**

| Elements    | Ti    | Zr    | Nb    | Mo    | Hf    |
|-------------|-------|-------|-------|-------|-------|
| Z           | 22    | 40    | 41    | 42    | 72    |
| r (Å)       | 1.46  | 1.60  | 1.43  | 1.36  | 1.58  |
| Atomic (%)  | 20.77 | 20.38 | 20.11 | 19.27 | 19.47 |
| Weight (%)  | 9.90  | 18.50 | 18.61 | 18.40 | 34.59 |